\documentclass[runningheads,a4paper]{llncs}

\usepackage[T1]{fontenc}

\usepackage{amssymb}
\usepackage{subfigure}
\usepackage{tabularx}
\usepackage{graphicx,wrapfig,lipsum}

\usepackage{epsfig}
\usepackage{subcaption}

\usepackage{url}
\urldef{\mailsa}\path|{alfred.hofmann, ursula.barth, ingrid.haas, frank.holzwarth,|
\urldef{\mailsb}\path|anna.kramer, leonie.kunz, christine.reiss, nicole.sator,|
\urldef{\mailsc}\path|erika.siebert-cole, peter.strasser, lncs}@springer.com|    
\newcommand{\keywords}[1]{\par\addvspace\baselineskip
\noindent\keywordname\enspace\ignorespaces#1}

\begin{document}

%
\title{Analyzing Cryptocurrency trends using Tweet Sentiment Data and User Meta-Data}

%
\author{Samyak Jain \inst{1} \and Sarthak Johari \inst{1}
\and Radhakrishnan Delhibabu \inst{2}}
\authorrunning{Samyak Jain et al.}
\titlerunning{Analyzing Cryptocurrency}

\institute{Computing Science and Engineering, Indraprastha Institute of Information Technology, New Delhi \and
School of Computing Science and Engineering, Vellore Institute of Technology,  Vellore
}

%
%
\toctitle{Lecture Notes in Computer Science}
\maketitle

\begin{abstract}
Cryptocurrency is a form of digital currency using cryptographic techniques in a decentralized system for secure peer-to-peer transactions. It is gaining much popularity over traditional methods of payments because it facilitates a very fast, easy and secure way of transactions. However, it is very volatile and is influenced by a range of factors, with social media being a major one. Thus, with over four billion active users of social media, we need to understand its influence on the crypto market and how it can lead to fluctuations in the values of these cryptocurrencies. In our work, we analyze the influence of activities on Twitter, in particular the sentiments of the tweets posted regarding cryptocurrencies and how it influences their prices. In addition, we also collect metadata related to tweets and users. We use all these features to also predict the price of cryptocurrency for which we use some regression-based models and an LSTM-based model.
\end{abstract}

\section{INTRODUCTION}
\subsection{MOTIVATION}
With the digitization of the world and the market, most of the financial operations are moving to the digital space. Cryptocurrency has emerged as a secure form of currency that allows end-to-end secured transactions. It has also emerged as a form of a financial asset just like traditional stocks in the stock market. 
However, the cryptocurrency exchange is an extremely volatile market that operates very differently compared to the traditional market. While traditional markets use technical indicators for calculating price fluctuations, the prices and valuations in cryptocurrency can be influenced by a wide range of factors ranging from the demand-supply balance, legal and regulatory factors, to the sentiments about it in news and social media. It has been clearly observed in the case of many popular virtual currencies that their prices can fluctuate heavily just on the basis of related activity on popular social media platforms like {\itshape Twitter, Facebook, etc}. 
Thus, our motivation for this research is to analyze the value fluctuations of cryptocurrencies from each bracket of market capitalization based on tweet sentiment analysis and use the user metadata for these tweets. We will analyze 2 coins from the large-cap like {\itshape Solana and Avalanche}, and 3 coins from the mid-cap range like {\itshape DogeCoin, Matic, and Shiba Inu}.

\begin{figure}[h]
  \centering
  \includegraphics[scale=0.8]{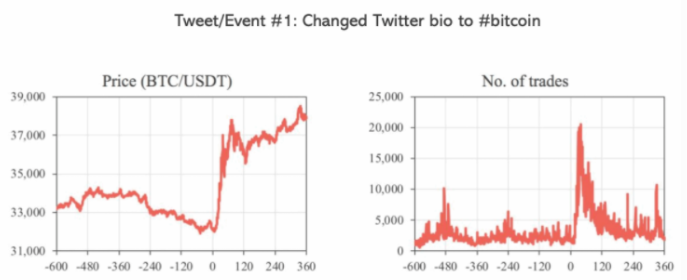}
  \caption{This figure shows the high fluctuation in bitcoin price and number of trades after Elon Musk, an influential person, changed his Twitter bio to \#bitcoin}
\end{figure}

\begin{figure}[h]
  \centering
  \includegraphics[width=\linewidth]{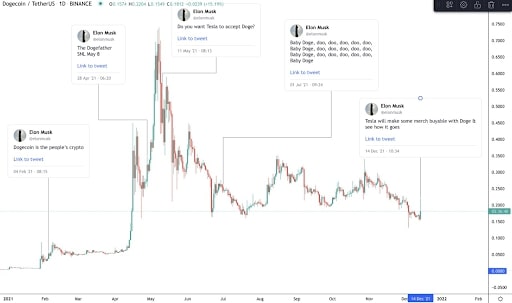}
  \caption{Trends of Dogecoin prices based on Twitter activity of Elon Musk, an influential person with a good reach.}
\end{figure}

\subsection{PROBLEM STATEMENT}
Our project aims to capture the sentiment of the text in the tweet to analyze the correlation between the price and sentiment of the cryptocurrency, and finally use that sentiment along with tweet metadata to conclude about the fluctuation in its price. We then use this sentiment coupled with user metadata as a combined feature to then predict the price of the cryptocurrency.

Our tasks include firstly collecting the relevant twitter data for the cryptocurrencies and the corresponding cryptocurrency values. Given a set of tweets that are related to a cryptocurrency coin, we would like to find the associated sentiment for that coin by performing sentiment analysis for the tweet text and then analyze the fluctuations in their value which occur in the near future as a result of these sentiments. We map the sentiment to a sentiment score ranging between zero and one which reflects the strength of the sentiment(positive/negative/neutral) and combine this with the tweet metadata to use the whole as a combined feature in our model for concluding about the fluctuation in the cryptocurrency’s value. Also, we would separately analyze this effect for both mid-cap range and high-cap range cryptocurrencies which differ in their market values.

The problem we are trying to solve is novel compared to the work previously done in this research area as we are not only focusing on the sentiment of the tweet but also incorporating user metadata and tweet data along with it to provide a deeper insight into user data and cryptocurrency fluctuation. The metadata includes details like count of followers, verification status, the number of likes, retweets etc. Including this as features in our models can provide better results which is because in terms of social media, the influence and effect created by the Twitter activity is very much dependent on the user who performs that as well (whether they are famous or influential or not). We are also trying to find out how high-cap range and mid-cap range coins are affected by the tweets and if the impact of a tweet is similar in both these cases.

\begin{figure}[h]
  \centering
  \includegraphics[width=\linewidth]{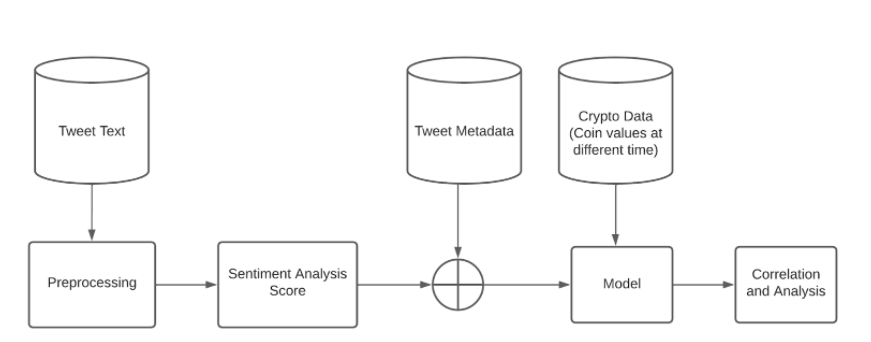}
  \caption{Project Pipeline}
\end{figure}

\subsection{MAJOR CONTRIBUTIONS}

Following are some major contributions of our work on this topic :
\begin{enumerate}
    \item We have collected the tweets and related data along with the price related data for two large-cap and three mid-cap cryptocurrencies. We ourselves have curated this dataset and released it for further use.

    \item We perform sentiment analysis of the tweet text to understand the context, and opinion by analyzing the polarity of the text. For this, we use RoBERTA based pre-trained model and finetuned the same. We tried to also unfreeze some layers of the already pretrained model while finetuning and then use the model to generate sentiment scores that are further used for our prediction and analysis.

    \item We analyze the effect of the sentiment of the tweets on the corresponding cryptocurrency price.

    \item We also use the sentiment analysis based features along with collected metadata related to the tweet and user to perform the task of price prediction for these cryptocurrencies. For this, we use regression-based models (linear regression, SGD regression and Random Forest regressor) and also to include a sense of time and sequence in our model we use LSTM to model this task as a time-series forecasting.

    \item We have tried out various approaches to train the LSTM based models with and without metadata. Also, we have accommodate the metadata after calculating the sentiment for a certain period of time and then weighing the sentiment by the number of retweets and likes based so as to capture the variation of sentiment and also metadata.
\end{enumerate}

\section{Literature Review}

[1] The paper predicts changes in Bitcoin and Ethereum (the two largest cryptocurrencies) prices using Twitter  data and Google trends. As Twitter is used widely as a news source and judging popularity, they influence the purchase/sell decisions of the user. They used three models to find the correlation with the cryptocurrency's price.  First, they collected tweets from twitter’s API using tweepy. After cleaning the collected tweets, they analyzed using the VADER\\(Valence Aware Dictionary for sEntiment Reasoning) sentiment analysis. Then they analyzed if the tweets actually have a sentiment or not and then established a relation between the sentiments of tweets with the price change of cryptocurrency. They found a positive correlation of prices with the sentiments when the price was rising. To have a better model input, they also considered the tweet volumes and used it as a metric to see the price fluctuation. They concluded that the relationship is robust to periods of high variance and non-linearity. With these inputs, a multiple linear regression model accurately reflected future price changes with the addition of lagged variables.

[4] This paper explores the use of social sentiment data as a better predictor as compared to the traditional methods of using technical financial indicators and uses the non-linear relation between sentiments and bitcoin price to predict prices in the future.\\
 
 \textbf{TRMI index construction} \\
        
    For this research, they have used an index called as Thomson Reuters Marketpsych Index (TRMI). It is evaluated on news, social media, and a combination of both. TRMI is defined as the ratio of the sum of all relevant variables to the sum of absolute value sum of the TRMI constituent variables, which is defined as Buzz. These are given as :
    \\
    \begin{equation}
    Buzz(a) = \sum_{c \in C(a), v\in V} |Var_{c,v}|
    \end{equation}
    \begin{equation}
    TRMI_t(a) = \frac{\sum_{c \in C(a), v\in V(t)} (I(t, v) * PsychVar_v(C))}{Buzz(Asset)}
    \end{equation}
     For the features obtained, they use ARIMA(Autoregressive Integrationg Moving Averages) and RNN  models. ARIMA has parameters such as autoregression, moving average, and integration. They also use variations of ARIMA such as ARIMAX which has an exogenous variable along with a time series variable attached to it. RNN is an artificial neural network-based model which takes the current input data and also the previous input data for making a prediction, which allows it to perceive data at time t-1. The paper shows that sentiment analysis is a key part of data-enabled algorithmic systems for cryptocurrency investments and trading.
    
[5] This paper uses historical tweet data fetched from twitter, user meta data containing the number of followers, number of retweets of a given tweet and corresponding Bitcoin prices, XRP prices and Ethereum prices at that given time instance because of the high correlation between their prices.
    The paper broadly works up on two aspects:
    \\
    \begin{enumerate}
        \item Implementing a predictive model which uses momentum metric to predict actions of buying, selling and holding a given crypto currency. If the momentum is above 5\% they predict buying, if less than -5\% they predict selling and if in between they predict holding. The threshold is decided based on the volatility of the crypto markets.
        \begin{equation}
          momentum = \frac{Price_{close} - Price_{open}}{Price_{open}}
        \end{equation}
        
        \item Providing explainability on the above implementation by using unsupervised deep learning clustering models to determine the underlying patterns. More formally, the paper compares each tweet representation obtained from DistilBERT to the buy, sell and hold category obtained from the section 1) and then finds the maximum similarity between groups of tweets and assigns the given tweet to the group having highest cosine similarity.

    \end{enumerate}
[6] The paper aims to study the correlation between Twitter sentiment and the changes it can bring to the prices of cryptocurrencies such as Bitcoin. Their motivation was that their research could help predict the price of Bitcoin in the future using past sentiments and Bitcoin prices. They wanted to develop a time series analysis for which tweets and Bitcoin prices along with their timestamps were collected from 12th March 2018 to 12th May 2018 as there was aggressive fluctuation on the prices of Bitcoin and this helped them in making an effective model. The data was scraped using APIs and web scraping techniques. Bitcoin prices were collected from four different sources namely - "BITSTAMP" "COINBASE",  "ITBIT" and  "KRAKEN'' and only close price of Bitcoin was used. In order to get a curve that is smooth the average of prices from the four sources was taken. Sentiment analysis using VADER (Valence Aware Dictionary and Sentiment Reasoner ) was performed. VADER was chosen as it was highly used when dealing with data from social media sites. A score between -1 to 1 was given by VADER. Finally, a Random Forest Regressor was used to perform evaluations of their model, Random Forest Regression was used as it was more adaptable to inputs of various kinds. A 62.48\% accuracy was observed while making the predictions using tweet sentiments and past prices of Bitcoin.

[7] The paper covers the study of Twitter sentiment about a particular altcoin (NEO) and how it correlates with its price in the crypto exchange market. They follow a straightforward methodology in their paper. The Twitter tweet data related to this particular altcoin is collected by scraping using fourteen different variations of a related hashtag for e.g - \#neo, \$neo, \#NEO etc. The tweet text along with some other tweet related data like username, language is collected using this way. This was followed by various preprocessing steps which included filtering based on the most frequent word, negative/positive crypto terms and removal of punctuation and spaces. Also, bot account tweets were also identified and filtered based of the frequency of tweets. Using a Random Forest classifier they were able to obtain 82\%  train set and 77\% test set accuracy for the sentiment analysis task on a subset. The pre-trained BERT in comparison had an accuracy of 45\% in the test set. The tweets' sentiments are classified into three classes: positive, negative and neutral. This sentiment is aggregated for a particular day and combined with cryptocurrency price and volume data(particularly bitcoin, Ethereum and NEO). The sentiment correlation with the price is observed and it was found that tweets with neutral sentiment had the highest correlation with NEO prices. Also, a high correlation was observed between Bitcoin and NEO prices. BERT based sentiment analysis showed that positive sentiment tweets had the highest correlation with price. Basically because neutral sentiment class is the most dominating, it was concluded that that is why probably it corresponded to a high correlation.\\
   
[3] The paper aims to predict the two-hour price  of cryptocurrency, namely Bitcoin and Litecoin on the basis of social factors such as tweet sentiments with the help of a multi-linear regression model. Bitcoin and Litecoin were chosen in particular due to their heavy popularity and reach among the public.The prices of the two bitcoins are extracted with the help of CoinDesk and the tweets are extracted with the help of rest APIs. After the tweets have been extracted they are classified into 3 classes on the basis of sentiments. Those sentiments being positive, negative and neutral. Textblob sentiment polarity is used for this purpose. This gives a score to a tweet between -1 and 1. All tweets with polarity $>$ 0 are classified as positive, all with polarity = 0 are classified as neutral and all tweets with polarity $<$ 0 are classified as negative. After this all these tweets are put into different groups based on the time they were posted and each group contains tweets posted within a span of 2 hours, the count of positive, negative and neutral tweets are kept as features. The average price during these 2 hours of both bitcoin and litecoin is also calculated and used as labels of the dataset. In the next phase, real-time tweets are used for testing and their metrics, accuracy and R2-score are used for the evaluation.

The proposed multilinear regression model is able to predict the 2-hour prices of bitcoin and litecoin upto an R2 value of 44\% and 59\% respectively. 

\section{METHODOLOGY}
\subsection{DATASET}

We extracted the data for six different cryptocurrencies, 3 coins from the large-cap range: Avalanche, Ripple, Solana and 3 coins from the mid-cap range: DogeCoin, Matic and Shiba Inu. For extracting the Twitter data, the Tweepy library is used which makes access of the Twitter API. We search for hashtags containing the symbol of the coin (‘\#<Name/Symbol of Coin>’) for searching the tweets relevant for the coin. The cryptocurrency price data for these coins are collected using the CryptoCompare API which provides historical cryptocurrency price data by minute, hour and day. We collected the cryptocurrency price data by the minute i.e at the interval of one minute.\\

\textbf{Fields in the tweet data include:}
\begin{itemize}
    \item id: tweet id
    \item text: tweet text
    \item favourite\_count:  The number of times the tweet has been favourited (liked).
    \item retweet\_count: The number of times the tweet has been retweeted.
    \item created\_at: The datetime of the moment the tweet has been tweeted.
    \item User: User related data for the user that tweeted it. It includes around sixty user-related information fields like id, name, screen\_name, location, followers\_count,\\ friends\_count, favourites\_count, verification status, following\_count etc.
    \item place: The place (geographical location) from where the tweet is tweeted.\\
\end{itemize}
\textbf{Fields collected in the cryptocurrency price data include:}
\begin{itemize}
    \item time: The datetime for which the crypto data is recorded
    \item high: The highest price during that time period (here minute)
    \item low: The lowest price during that time period (here minute)
    \item open: The price at the start of the minute
    \item volume\_from: Total amount of base currency (USD) traded into the cryptocurrency during that minute.
    \item volume\_to: Total amount of cryptocurrency traded into the base currency (USD) during that minute.
    \item close: The price at the end of the minute\\
\end{itemize}

\subsubsection{Preprocessing}
The collected Twitter data included the tweet text which was preprocessed because it is needed for the part of sentiment analysis. Also, the ‘User’ data in the tweet data was present in the JSON form and the keys of the JSON were parsed into columns of our pandas dataframe. The preprocessing steps applied to the tweet text include:

\begin{itemize}
    \item Removed any user mentions present and handled retweets: Tweets tend to have user mentions “@sarthakj01” such mentions are removed from the tweets
    \item Removed any links/URLs from the tweet text: Tweets have links, all HTTP or bitly links are removed from the tweets 
    \item Separating the hashtags in a different column, these hashtags are useful as they contain very important information regarding the tweet: Hashtags such as \# bitcoin,  \# DOGECOIN are handled, they are removed from the tweets and kept in a separate column for that tweet as they can hold vital information
    \item Converted tweets to lowercase
    \item Removed punctuation
    \item Classified Emojis and Emoticons: Emoticons like :-)) is written as ‘Very\_happy’ and symbol based emojis are also appropriately expanded. Python’s emot and emoji libraries are used for these respective classifications\\
\end{itemize}    

\textbf{Outlier Detection}\\

\begin{itemize}
    \item In the Avalanche dataset, we found out that there were many tweets which contained the word Avalanche but weren't related to the cryptocurrency Avalanche
    \item Upon further research we found out that there were multiple other meanings related to the word Avalanche (such as a football team) and hence the tweets were out of context
    \item For this problem we created a list of crypto/financial terms and classified the data into crypto or non crypto category and discarded the non crypto category data because that would have led to wrong results while sentiment and further analysis\\
\end{itemize}

Note that we deliberately chose not to remove the stopwords because in the case of sentiment analysis they can hold important sentiment-related information.  Removing them can lead to capturing the sentiment wrongly.\\For example:
\\\\\textbf{Original sentence:} Bitcoin is not a good investment (Negative sentiment)\\
\textbf{After stopword removal:} Bitcoin good investment (Positive sentiment)\\

The cyrptocurrency price data did not need any preprocessing. The cryptocurrency price data is finally joined with the tweets data on the basis of the time of the tweet. The cryptocurrency price data of the very next minute of the time given by the  ‘created\_at’ column of the tweet is joined with that tweet’s data. A tweet made at the time 
‘hh:mm:ss’ will have the price data of that cryptocurrency at ‘hh:(mm+1):00’ joined with it.

\subsubsection{Analysis}
\begin{itemize}
    \item We generate the wordcloud of our twitter dataset for each coin. The tweet text was first preprocessed according to the preprocessing steps explained above. (Figure ~\ref{fig:wordcloudforcoins}) 
    
    \begin{figure}[h]
      \centering
      \includegraphics[scale=0.4]{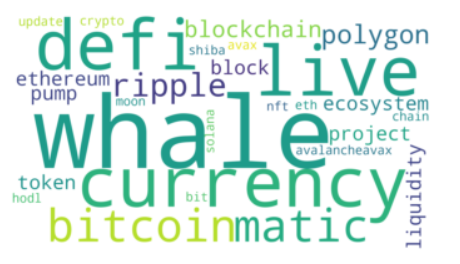}
      \caption{This figure shows the most occurring words in the crypto context}
     
      \label{fig:cryptoconextwords}
    \end{figure}

\begin{figure}
  \begin{minipage}[b]{0.5\linewidth}
    \centering
    \includegraphics[width=1\textwidth]{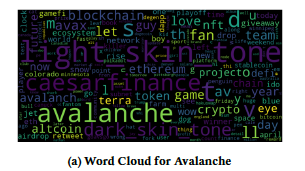}
  \end{minipage}
  \hfill
  \begin{minipage}[b]{0.5\linewidth}
    \centering
    \includegraphics[width=0.9\textwidth]{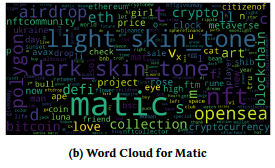}
  \end{minipage}
  \caption{Word clouds}
\label{fig:wordcloudforcoins}
\end{figure}
 
     \item   Then we went for topic modelling of our tweet text. For this we used pyLDAvis library which gives a visual representation to the topic modelling performed by Latent Dirichlet Allocation. This helps us to understand about the most popular topics in our dataset, The preprocessed text is used for LDA.
    
    \item To understand if the fact of a user being verified have any impact to the way people view a tweet, we plotted a graph (Figure ~\ref{fig:userverified}) , it can be clearly seen that people engage more with a tweet when the user is verified hence there is a much greater retweet count and favourite count when compared to the tweet made by a user who is not verified.
    \item The average tweet text length is also calculated for all the different coins' tweet data (Figure ~\ref{fig:tweetlens}).
    \begin{figure}[h]
      \centering
      \includegraphics[scale=0.5]{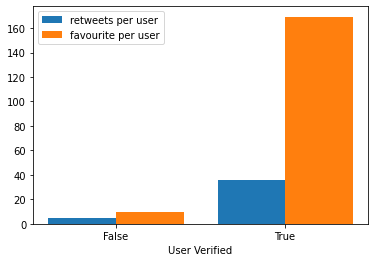}
      \caption{This figure shows the total retweet count and favourite count when the tweets are made by verified users and when the tweets are made by unverified users}
      \label{fig:userverified}
    
    \end{figure}
    
    \begin{figure}[h]
      \centering
      \includegraphics[scale=0.5]{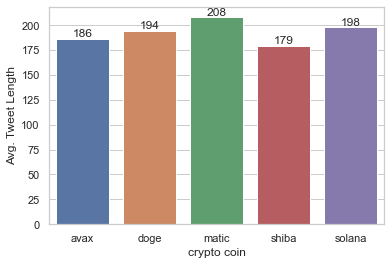}
      \caption{Average Tweet Length for all coins}
      \label{fig:tweetlens}
    
    \end{figure}
\end{itemize}

\subsection{Sentiment Analysis}

\begin{itemize}
    \item Sentiment Analysis as a NLP task where we identify, and categorize opinions expressed in a given text, with respect to the overall sentiment of the corpus. This means that for a review based dataset, a positive tweet will have a different emotion attached to it, as compared to a finance based dataset.
    
    \item For our problem statement the preprocessed dataset has an undefined sentiment score. Using a pretrained sentiment analysis model might not lead to accurate results for the same reasons mentioned above. Hence, we decided to fine tune a pretrained sentiment analysis model using a previously collected and labeled dataset for crypto currency tweets. This will further improve the sentiment analysis results.
    \item To this extent, we have used a BitCoin dataset, with tweets and their sentiment category (\emph{positive, neutral, negative}) and a related sentiment score. The given dataset had a total of 50,859 tweets. Out of these tweets, we use 24917 tweets for training, 15257 tweets for testing, and 10679 tweets for validation.
    \item \emph{Pre-Trained Model}: For our pre-trained model, we use the \emph{twitter-roberta-base-sentiment} model, which was previously trained on around 58M tweets, and fine tuned for sentiment analysis with the TweetEval benchmark.
    \item \emph{RoBERTa} : A futher improvement on BERT model to improve pre-training in self-supervised NLP systems. RoBERTa improves on BERT by training larger mini-batches, improved learning rates, and removing next-sentence pre-training object in a BERT model.
    \item For training the given model, we use a learning rate of 1e(-5), batch size of 8 and a total of 4 epochs. The maximum length is set to 256, which is under the average tweet text length we have. After fine tuning the model, we use it on our current dataset.
    \item In the predictions obtained from our model, we get a sentiment label and the corresponding sentiment score which is a value in the range [0,1] on the tweet text of the tweet dataset we have for all the coins.

     \item We used various models which are BERT, Roberta based and available on hugging face, we finally used the model which was Roberta based and pretrained on twitter sentiment analysis and finetuned it on our crypto dataset by splitting it into train and validation sets as it gave the best accuracy
     \item Finally we obtained our predictions which are the sentiment label (Positive, Negative, or Neutral) and the corresponding sentiment score which is a value in the range [0,1] on the dataset we had obtained by scraping tweet data and computing sentiment scores for cryptocurrencies such as Avalanche, Doge Coin, Matic, Solana and Shiba Inu.
\end{itemize}

\subsection{Prediction Models}

We use regression based machine learning models from scikit-learn to make a prediction on crypto coin prices. These models use the sentiment scores and sentiment labels of the tweet text along with tweet metadata(favourite\_count, retweet\_count) and basic user metadeta (user\_follower\_count, user\_verified) as feature set. The columns of sentiment label is one-hot encoded and user\_verified status is mapped to a binary integer value (0,1).\\\\
Regression models used for baseline price prediction include:\\
\begin{itemize}
    \item \textbf{Linear Regression (LR)}: It tries to fit a linear model with the help of coefficients $W = (W_1, W_2, W_3, W_4, ….., W_n)$ so as to reduce the residual sum of squares between the actual values and the values predicted using the linear approximation.\\
    \item \textbf{SGD Regressor (SGD-R)}: It works towards building an estimator using a regularized linear model. The regularizer adds a penalty to the loss to help shrink the model parameters. It follows a stochastic gradient descent method where gradients are computed for each sample one at a time and weights are updated using that. We use Huber loss during training of the model. It is less sensitive to outliers and is computed as:\\

\begin{figure}
\centering
\includegraphics[scale=0.6]{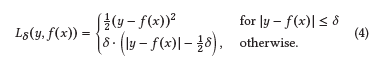}
\label{fig:example}
\end{figure}

 \item \textbf{Random Forest Regressor (RF-R)}: It is an ensemble method that tries to improve the predictive power by fitting a number of classifying decision trees and averaging.\\
\end{itemize}
We split our data into training and testing sets for running through the machine learning models.We have taken a 70:30 train:test split ratio.The parameters used for training SGDRegressor include 'l2' penalty, initial learning rate of 0.01, maximum\_iter as 100 and a alpha(regularization weight) of 0.01. We take the max\_depth of the Random Forest Regressor as 5 and the Linear Regression model is trained on default settings. The metrics we use for evaluation of the performance of these predictive models are Mean Absolute Error (MAE) and Root Mean Squared Error(RMSE). In addition to this, we also use a metric called percentage error ($\delta$) for comparisons.\\

\begin{figure}
\centering
\includegraphics[scale=0.6]{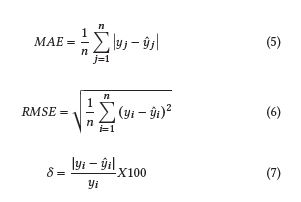}
\label{fig:example}
\end{figure}

\textbf{LSTM based model }: As the problem of price prediction has a time variable and also is related to the previous values and features in the real world, we have used LSTM based model. Although RNN models can theoretically remember about all previous occurrences but in practice that is not the case, the LSTM is an optimized version of RNNs that can in practice also remember the past occurrences and by a mechanism of hidden state, forget input and output gates. 

\begin{wrapfigure}{r}{5.5cm}
\includegraphics[width=5cm]{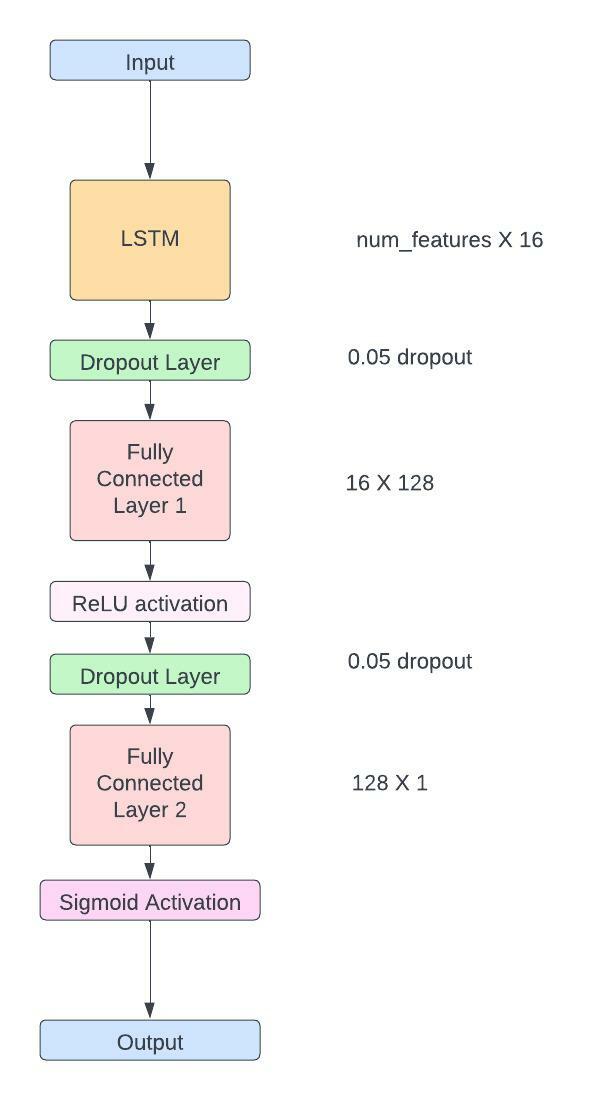}
\caption{LSTM based model architecture}\label{fig:modelarch}
\end{wrapfigure} 

We tried various stacked combinations and then used fully connected layers with combinations over it to finally obtain the predictions. We train for 200 epochs with a learning rate of 8e-4, hidden size 16. We also add price related features like high, low, volume\_to/from etc to the data. Also, we both standard scale the data and do min-max normalization on the price values while training. However, we also take appropriate inverse transforms while inference of the results whenever needed. The model architecture can be seen in Figure~\ref{fig:modelarch}\\.

\section{EVALUATION}
\medskip
We evaluate and analyze the result of the price prediction task and the sentiment analysis part for all the cryptocurrencies.
\subsection{Prediction Models}
The above mentioned linear models are trained on the features dataset and the corresponding cryptocurrency coin price is predicted. Following are the values of mean absolute error, root mean squared error and maximum percentage error obtained for each model:

\begin{figure}
  \begin{minipage}[b]{0.5\linewidth}
    \centering
    \includegraphics[width=1\textwidth]{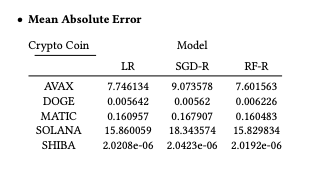}
  \end{minipage}
  \hfill
  \begin{minipage}[b]{.5\linewidth}
    \centering
    \includegraphics[width=1\textwidth]{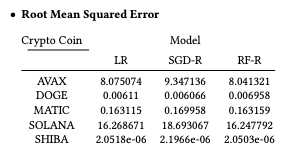}
  \end{minipage}
\end{figure}

From the loss values we observe that learning is happening the best for ShibaInu, DogeCoin and Matic. Their predicted values are close to the actual values. Avalanche and Solana have high error/loss value which indicates the model is not learning to predict their prices properly. An important thing to note here is that the price data is time dependent i.e. the price of the cryptocurrency coin is dependent on the previous prices and the previous feature data (here sentiment data) as well. Thus here in our regression-based baselines we have not considered this information (dependency on past data and values) and there is no context of time or sequence in our current models which work considering the features as simple numerical data. This leads to loss of contextual information.

\begin{figure}
  \begin{minipage}[b]{0.5\linewidth}
    \centering
    \includegraphics[width=1\textwidth]{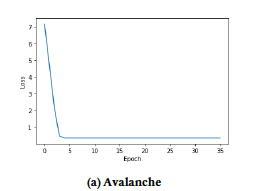}
  \end{minipage}
  \hfill
  \begin{minipage}[b]{.5\linewidth}
    \centering
    \includegraphics[width=1\textwidth]{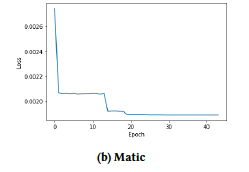}
  \end{minipage}
\caption{SGD Regressor training loss Curves}
\end{figure}

Following are the values of mean absolute error, root mean squared error and maximum percentage error (max $\delta$) obtained for LSTM based model:

\begin{figure}
  \begin{minipage}[b]{0.5\linewidth}
    \centering
    \includegraphics[width=1\textwidth]{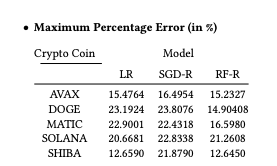}
  \end{minipage}
  \hfill
  \begin{minipage}[b]{.5\linewidth}
    \centering
    \includegraphics[width=1\textwidth]{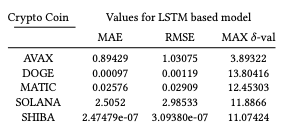}
  \end{minipage}
\end{figure}

\begin{wrapfigure}{r}{3.5cm}
\includegraphics[width=0.35\textwidth]{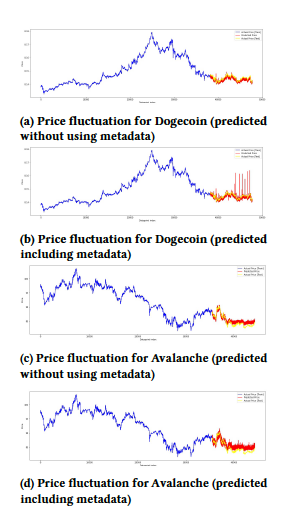}
\caption{Price value over time for Dogecoin and Avalanche}
\label{fig:modelpriceplot}
\end{wrapfigure}

We observe a significant improvement in the MAE and RMSE values when compared with baseline models. Most significantly, we can see an improvement in the value of maximum percentage error for all the crypto coins. Also, when we plot all the price values of the cryptocurrency (both training set and predicted values along with actual values), we can see that the model is capturing the trend of the prices and is predicting values in a good range. This holds true for both Avalanche and Solana as well that had high MAE and RMSE values in the regression based models. The fluctuations in the prices are also being properly captured. Figure ~\ref{fig:modelpriceplot} depicts this for two of the crypto coins. Also, on analyzing the effect of metadata on the predictions, we find that the error values remain almost similar but introducing metadata adds a bit of noise to the predictions i.e- we observe that there are some spikes in the predictions.

Since, we do not find any other works that use the same large-cap and mid-cap coins as used by us, we cannot make a direct comparison for our results. However, the recent work on bitcoin price prediction using twitter sentiment analysis reports a maximum percentage error of 43.83\% [6]. The actual and predicted price value's fluctuation for Bitcoin and its plot is also presented in this work [6].Another work on bitcoin price prediction report the best MAE value of 2.7526 and RMSE value 13.7033 [2].  Comparing to the cryptocurrencies we have chosen, we can see that we achieve significantly better(lower) MAE and RMSE value for all of them in our best case LSTM based model. Also, we have a notably lower value of the maximum percentage error than these. Also, our price plots are more smoother and have lesser noise and capture the tend more accurately.

\subsection{Sentiment Analysis}
\begin{figure}
  \centering
  \includegraphics[width=\linewidth]{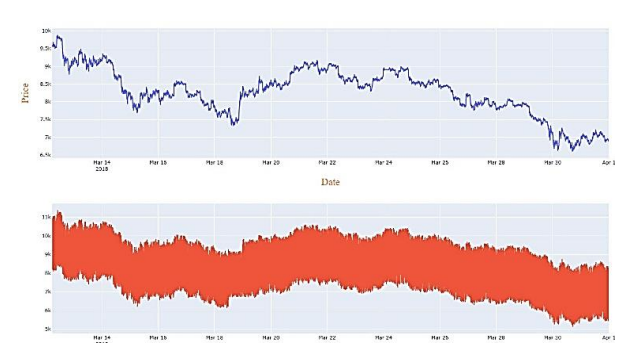}
  \caption{Actual Price Plot and Predicted Price Plot for Bitcoin as presented in [6]}
  \label{fig:papercrypto}
\end{figure}
\textbf{Principal Component Analysis} : After evaluating our dataset on the fine tuned sentiment analysis model, we find the sentiment score and sentiment label for each tweet, for all the coins.  However, we need to analyze whether the text and their corresponding sentiments are actually trained well or not. To do this we can use clustering to find whether similar sentiments are clustered together. For this, we use PCA, a dimensionality reduction technique. First, we use a sentence tokenizer(BERT based) on all the tweets, and pick up 1000 points randomly from each of the sentiment classes. This results in an embedded matrix of size 3000 X 786 (no. of tweets X embedding size). We  apply PCA, reducing their total dimensions to 2. Using these dimensions we plot the points, and observe that similar sentiments get clustered together. 

\begin{figure}[h]
  \begin{minipage}[b]{0.5\linewidth}
    \centering
    \includegraphics[width=1.2\textwidth]{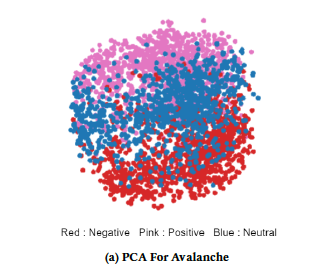}
  \end{minipage}
  \hfill
  \begin{minipage}[b]{.5\linewidth}
    \centering
    \includegraphics[width=1\textwidth]{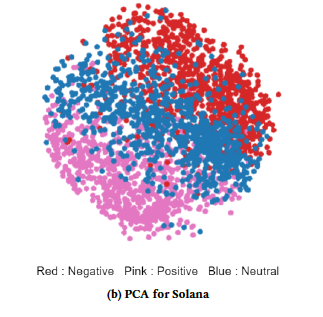}
  \end{minipage}
\caption{PCA for crypto sentiments}
          \label{fig:pcafigs}
\end{figure}

\begin{figure}[h]
  \begin{minipage}[b]{0.5\linewidth}
    \centering
    \includegraphics[width=1.2\textwidth]{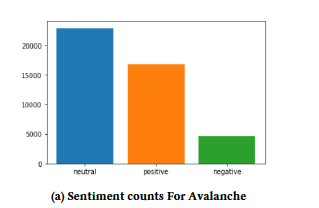}
  \end{minipage}
  \hfill
  \begin{minipage}[b]{.5\linewidth}
    \centering
    \includegraphics[width=1\textwidth]{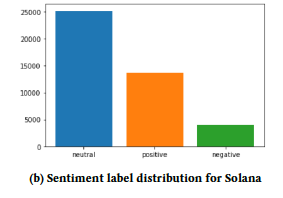}
  \end{minipage}
\caption{Sentiment label distribution for crypto}
          \label{fig:sentimentcountsfigs}
\end{figure}

\begin{figure}[h]
  \begin{minipage}[b]{0.5\linewidth}
    \centering
    \includegraphics[width=1\textwidth]{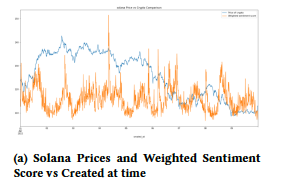}
  \end{minipage}
  \hfill
  \begin{minipage}[b]{.5\linewidth}
    \centering
    \includegraphics[width=1\textwidth]{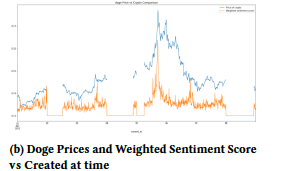}
  \end{minipage}
\end{figure}

\begin{figure}[h]
  \begin{minipage}[b]{0.5\linewidth}
    \centering
    \includegraphics[width=1\textwidth]{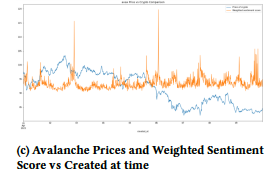}
  \end{minipage}
  \hfill
  \begin{minipage}[b]{.5\linewidth}
    \centering
    \includegraphics[width=1\textwidth]{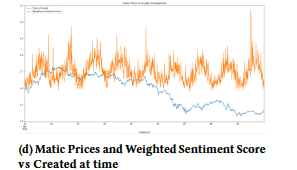}
  \end{minipage}
\end{figure}

\begin{figure}
  \centering
  \includegraphics[scale=0.5]{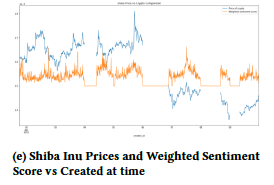}
\caption{Crypto Prices and Weighted Sentiment Score vs Created at time}
\label{fig:weightedsentimentscore}
\end{figure}

We plotted prices of crypto currency and the weighted sentiment score against created\_at time of the tweet (Figure ~\ref{fig:weightedsentimentscore} ). As we hoped, we were able to find some correlation in between the weighted sentiment score. It could be seen from the graph of almost all the currencies that the peaks and declines for both weighted sentiment score and the crypto currency coincided with time, thus showing that positive sentiments have led to an increase in the prices of cryptocurrency and and the negative sentiments have led to a decline. We also not that they do not coincide at the exact same moment rather, the effect of sentiment on price of the cryptocurrency comes after some hours. This causal relation is visible more clearly for DogeCoin and Avalanche.
The weighted sentiment score is calculated by assigning certain weights to positive, negative and neutral sentiment labels and then scaling this according to the cryptocurrency price (so that the sentiment value and prices are in comparable range for plotting). The weights are decided empirically for all coins.

\section{CONCLUSION}
We have computed the sentiment score and label of the collected tweets for the 2 large-cap crypto coins Solana and Avalanche, and 3 coins from mid-cap range like Dogecoin, Matic and Shiba Inu. It is done using RoBERTa based pretrained sentiment analysis model finetuned on a crypto-sentiment (Bitcoin-based) dataset. We observe that similar sentiments get clustered together. We combine this sentiment analysis derived features with the metadata we collect which include features like favourite\_count, retweet\_count and user metadata (follower\_count, user\_verified). 

We have used this for the task of price prediction and use three regression based models and an LSTM based model. We obtain satisfactory results in terms of the metrics like MAE, RMSE and percentage error. We also see from the price plots (Figure 10) that the model is able to capture the trend of price change of the cryptocurrency and is appropriately predicting its future value.

Now moving forward things that can be worked upon more are that our model predicts a decently well accuracy but the predictions are shaky, that is the model is not able to capture well that the price fluctuations are almost continuous in nature. More complex models might be able to capture this and as the duration of our project was less and more time was occupied by other things we could not deep dive into more complex model architectures that might be able to capture this well.

Also some mechanism to incorporate the sentiment and corresponding tweet metadata can be tinkered with as the tweets seem noisy. This can be because of the twitter bots tweeting similar kind of posts that makes the sentiment inflated to either the positive or negative side. Some more advanced and complex time series forecasting methods can also be used for this task in the future.

\end{document}